\documentclass[aps,prl,reprint,showpacs,showkeys,superscriptaddress]{revtex4-1}
\usepackage{amsmath,amssymb,bm,mathrsfs}
\usepackage{graphicx}
\usepackage{srcltx}
\usepackage{hyperref}

\begin{document}
\title{Low-energy Supersymmetry Breaking Without the Gravitino Problem} 

\author{Anson Hook}
\email{hook@ias.edu}
\affiliation{School of Natural Sciences, Institute for Advanced Study, Einstein Drive, Princeton, New Jersey 08540 USA}

\author{Hitoshi Murayama}
\email{hitoshi@berkeley.edu, hitoshi.murayama@ipmu.jp}
\affiliation{Department of Physics, University of California,
  Berkeley, California 94720, USA} 
\affiliation{Theoretical Physics Group, Lawrence Berkeley National
  Laboratory, Berkeley, California 94720, USA} 
\affiliation{Kavli Institute for the Physics and Mathematics of the
  Universe (WPI), Todai Institutes for Advanced Study, University of Tokyo,
  Kashiwa 277-8583, Japan} 

\begin{abstract}
  In models of low-energy gauge mediation, the observed Higgs mass is
  in tension with the cosmological limit on the gravitino mass
  $m_{3/2} \lesssim 16$~eV.  We present an alternative mediation
  mechanism of supersymmetry breaking via a $U(1)$ $D$-term with an
  $E_6$-inspired particle content, which we call {\it vector
    mediation}\/.  The gravitino mass can be in the eV range.  The
  sfermion masses are at the 10~TeV scale, while gauginos around a
  TeV.  This mechanism also greatly  ameliorates the $\mu$-problem.
\end{abstract}

\maketitle

\paragraph{Introduction.}

Supersymmetry (SUSY) is the prime candidate for solving the
naturalness problem of the Standard Model Higgs sector (see, {\it
  e.g.}\/, \cite{Murayama:2000dw}).  
Models of low-energy SUSY breaking are particularly attractive because
the light gravitino $\tilde{G}$ can also be produced at colliders with
the striking signature of additional photons in the decay of the bino
$\tilde{B} \rightarrow \tilde{G} \gamma$ \cite{Cabibbo:1981er}.  The
most extensively studied mechanism for this is gauge mediation
\cite{AlvarezGaume:1981wy}.

However, cosmological limits on the gravitino are in strong tension
with low-energy gauge mediation.  In the Minimal Supesymmetric
Standard Model (MSSM), the observed Higgs mass of
125~GeV~\cite{Aad:2012tfa,Chatrchyan:2012ufa} requires large $A$-terms
or stop masses in the 10~TeV range.  In gauge mediation, generating
large $A$-terms through renormalization group
evolution~\cite{Draper:2011aa} or large stop masses requires the SUSY
breaking scale to be above 1000~TeV, and hence the gravitino mass
$m_{3/2}$ to be above a keV~\cite{Yanagida:2012ef}.  Such a gravitino
would overclose the Universe unless the reheating temperature after
the inflation is kept unnaturally low~\cite{Moroi:1993mb,
  deGouvea:1997tn}, leaving little room for any conceivable
baryogenesis scenario.  When $m_{3/2} \lesssim 1$~keV, the gravitino
does not overclose the Universe, but would be a warm dark matter
component, ruled out by the Lyman-$\alpha$ forest data unless $m_{3/2}
\lesssim 16$~eV~\cite{Viel:2005qj}.  The tension may be eased if the
Higgs sector is extended to the NMSSM or Dirac--NMSSM~\cite{Lu:2013cta},
if the messenger sector couples to the Higgs~\cite{Craig:2012xp}, or
is strongly coupled~\cite{Yanagida:2010zz,Heckman:2011bb,
  Evans:2012uf,Kitano:2012wv}.  However, these models are somewhat
more complicated.

In this Letter, we present an alternative simple mediation mechanism of
SUSY breaking, where the sfermion masses are generated at
tree-level from a $U(1)$ $D$-term.  $m_{3/2}$ can be brought down to
the eV scale, solving the cosmological problem.  In addition, the
gluino is most likely within the reach of the HL-LHC.

\paragraph{Vector Mediation}

We first present the basic idea of what we call the {\it vector
  mediation}\/ of SUSY breaking.

We assume that the SUSY breaking sector produces both an $F$-term and
a $D$-term of a $U(1)$ vector multiplet under which the Standard Model
matter fields are charged.  In order for all of the sfermions in the
Standard Model to acquire positive soft mass-squared, all quark and
lepton multiplets must have $U(1)$ charges of the same sign (positive).
Anomaly cancellation requires additional superfields with
negative $U(1)$ charges, which acquire negative soft
mass-squared from the $D$-term.  They must come in Standard Model
vector-like representations so that the vector-like masses can
overcome the negative soft mass-squared.  It is economical if the
vector-like supermultiplets also act as messengers that generate
gaugino masses at the one-loop level.  Because of the tree-level
generation of scalar masses, the scalar masses as well as the scale of
SUSY breaking are both in the 10--100~TeV range, solving the
cosmological gravitino problem.

The tree-level mediation of SUSY breaking via a $U(1)$
$D$-term was also considered in
Refs.~\cite{Nardecchia:2009ew,Monaco:2011fe}.  Their interest was in a
high-scale mediation mechanism which has a much more weakly coupled
goldstino and thus have very different cosmology and collider
signatures~\cite{Arcadi:2011yw}.

\paragraph{An Explicit Model}

The simplest particle content that satisfies the requirements
explained above is one inspired by $E_6$ unification.  By
embedding $SO(10) \times U(1)_\psi$ into $E_6$, the fundamental
representation ${\bf 27}$ decomposes as
\begin{equation}
  {\bf 27} = \Psi ({\bf 16}, +1) \oplus \Phi ({\bf 10}, -2) \oplus S
  ({\bf 1}, +4).
\end{equation}
We identify $\Psi$ as the quark and lepton supermultiplets, while
$\Phi$ plays the role of messengers as well as the Higgs.  $S$ is
responsible for the SUSY breaking.

\paragraph{SUSY Breaking}

We introduce a vector-like mutiplet $X(-4)$ and $Y(+4)$ charged under
$U(1)_\psi$.  We also introduce a neutral field $Z(0)$ and write the
superpotential 
\begin{equation}
  W = M S X + \lambda Z (X Y - v^2).
\end{equation}
It is essential that the model here is {\it chiral}\/ under
$U(1)_\psi$, since charge conjugation invariance would prevent the
generation of a $D$-term.  Note that the above superpotential is
generic and $U(1)_R$ invariant, which guarantees that it breaks
SUSY~\cite{Nelson:1993nf} \footnote{We will not impose the exact
  $U(1)_R$ symmetry on the whole theory.  We simply assume that
  $U(1)_R$ breaking terms such as $Z^2$, $Z^3$ are sufficiently small
  so as not to destabilize the desired vacuum.}.

This model breaks SUSY with a positive $D$-term, which
induces tree-level positive mass-squared to the matter fields $\Psi$.  
Motivated by models of dynamical SUSY breaking, {\it \`a la}\/
Izawa--Yanagida--Intriligator--Thomas
(IYIT)~\cite{Izawa:1996pk,Intriligator:1996pu}, we take the values
$\lambda = 4 \pi$, $e=0.1$ and $M=2v$ as an example.  We will use
the notation $X_0 = \langle X \rangle$, etc.  The minimum is at
$X_0=0.58v$, $Y_0 = 1.71v$, $S_0=0$ and $Z_0=0$.  The non-zero
SUSY breaking parameters are $F_S = M X_0 = 1.16v^2$, $F_Z =
\lambda(X_0Y_0-v^2) = -0.1v^2$, and $D = 4e(-|X_0|^2 + |Y_0|^2) =
1.02v^2>0$.  All particles here are massive except for the
goldstino, which is a linear combination of the $S$ fermion and the
$U(1)_\psi$ gaugino.  Hence there is no cosmological Polonyi
problem~\cite{Coughlan:1984yk}.


Note that the size of the $D$-term is completely fixed by the size of
the $F$-term as
\begin{equation}
  D = \frac{2}{M_V^2} F_S^* (4 e) F_S, \quad
  M_V^2 = 2 (4 e)^2 ( |X_0|^2 + |Y_0|^2 ).
\end{equation}
Here, $M_V$ is the mass of the $U(1)_\psi$ gauge boson.  This equation
is a simple consequence of the equation of motion for the massive
vector multiplet.

In our $E_6$ inspired model, there are two additional $S_{1,2}$ fields
which do not play a role in SUSY breaking.  To give $S_{1,2}$
a mass, we introduce two new neutral fields $N_{1,2}$ and introduce
the couplings $W = \kappa N_{1,2} S_{1,2} X$.  Once $X$ obtains its
vacuum expectation value, this superpotential combined with the
$D$-term gives a mass to $S_{1,2}$ and $N_{1,2}$.

\paragraph{Tunneling to the true vacuum}

In order to generate gaugino masses at one-loop, we need the superpotential
\begin{equation}
  W_\Phi = -\frac{g}{2} Y \Phi \Phi - \frac{k}{2} S \Phi \Phi. \label{Eq: messenger}
\end{equation}
Because this superpotential does not respect the $U(1)_R$ symmetry, a
supersymmetric vacuum appears.  It is located
at
\begin{equation}
  X_0 = \frac{v^2}{Y_0} = \frac{k \Phi^2_0}{2 M}, \quad 
  Z_0 = \frac{g M}{k \lambda}, \quad
  S_0 = -\frac{2 g M v^2}{k^2 \Phi^2_0},
\end{equation}
where the value of $\Phi_0$ is determined by minimizing the $D$-term.
Because there is both a SUSY-preserving and a SUSY-breaking
vacuum, one must consider tunneling between the two vacua.  From the
form of the supersymmetric vacuum, it is clear that if $k$ is small,
the distance between the two vacua will be large.  In the limit that
$k \rightarrow 0$, the SUSY vacuum disappears and the metastable
vacuum becomes stable.  

The life-time of the metastable vacuum can be calculated using
semiclassical techniques~\cite{Coleman:1977py}.  The tunneling rate
per unit volume is $\Gamma \propto e^{-B}$ where $B = S_E(\overline \phi(r)) 
- S_E(\phi_0)$.  $\phi_0$ is the field configuration of the metastable
vacuum, $S_E$ is the Euclidean action and $\overline \phi(r)$ is the bounce
profile.  The Euclidean action is
\begin{equation}
  S_E(\phi(r)) = 2 \pi^2 \int_0^\infty dr r^3 \left[ \sum_i \frac{1}{2} \
    \left(\frac{d \phi_i}{dr}\right)^2 + V(\phi(r)) \right],
\end{equation}
where the sum goes over all five fields, $\phi_i = \left ( X, Y, Z, \Phi, S \right )$.  
$\overline \phi(r)$ is the solution to
\begin{equation} 
\label{eq:vacuum bounce}
\frac{d^2 \phi_i}{dr^2} + \frac{3}{r} \frac{d \phi_i}{dr} = \frac{dV}{d\phi_i}\ ,
\end{equation}
subject to the boundary conditions 
\begin{equation}
\frac{d\overline \phi_i}{dr}(r=0) = 0, \qquad 
\overline \phi_i(r \rightarrow \infty) = \phi_0.
\label{eq:boundary}
\end{equation}
We need $B \gtrsim 450$ for
the metastable minimum to live longer than the age of the universe.


Calculating the bounce properly requires solving a five-dimensional
differential equation.  
To simplify the computation,
we approximate the tunneling as confined to a
one-dimensional sub-space proceeding in a straight line between the
two minima.  In this approach, solving for the
bounce action is numerically very stable.

We expect this approximation to be valid when the potential has large
first derivatives, {\it e.g.}\/, when the maximum separating the two
minima is very large.  Then the friction term, the $1/r$ term in
Eq.~\eqref{eq:vacuum bounce}, is important.  In order for the bounce
to travel between the two minimum, it first stays near the SUSY minima
until large $r$ where the friction term is irrelevant.  The bounce
then rolls to the SUSY breaking minimum.  In this limit, because the
bounce action is required to start very close to the SUSY minima and
to end at the SUSY breaking minimum, we expect that this
one-dimensional approximation is good.
We have also checked that the maximum separating the two minima
lies on the straight line between the two minima to within a few percent.

\begin{figure}[t]
  \centering
  \includegraphics[width=0.4\textwidth]{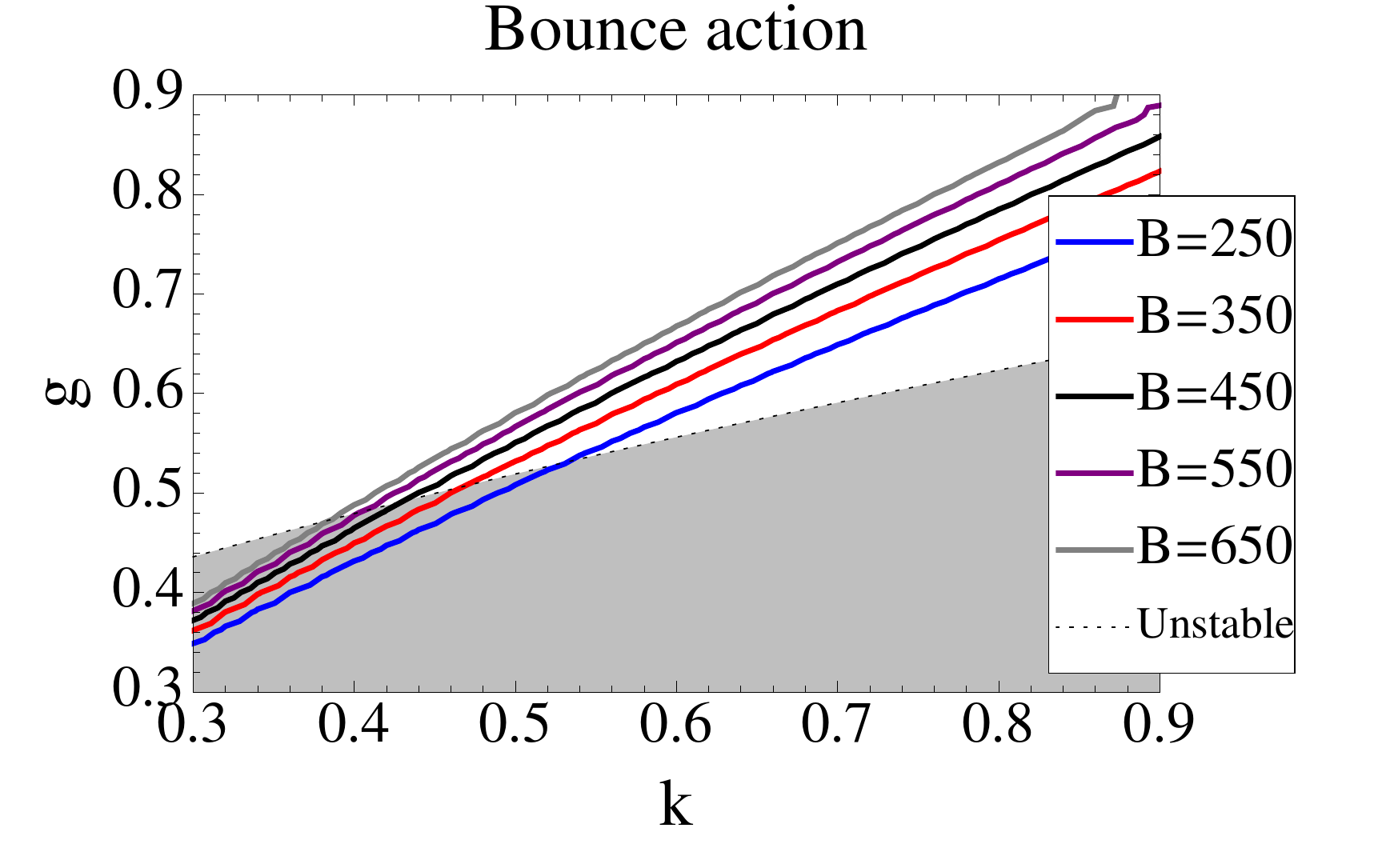}
  \caption{Contours of the vacuum bounce action on the $(k,g)$ plane.
    The solid blue, red, black, purple and gray lines are for
    $B=250$, 350, 450, 550, and 650, respectively.  For the universe
    to be stable, $B \gtrsim 450$.  The dashed black line and the
    grey region show where the SUSY breaking minimum becomes
    tachyonic.  We expect that the approximations used to calculate
    the bounce action fail in this region.}
  \label{fig:bounce}
\end{figure}

In Fig.~\ref{fig:bounce}, we plot contours of the bounce action on the
$(k,g)$ plane.  The shaded region is where the messengers go
tachyonic.  Then there is no maximum between the two vacua and the
bounce does not start near the SUSY vacua.  Thus, we expect the
approximation used to calculate the bounce action to be very poor.  We
have verified numerically that the one-dimensional approximations of
the bounces in these regions of parameter space do not start near the
SUSY vacua.

Aside from the vacuum transition rate, there is also the probability
of thermally tunneling between the two vacua in the early universe.
For $T \gtrsim v$, the SUSY vacuum disappears and there is only the
SUSY-breaking vacuum.  At a lower temperature, the SUSY-preserving
vacuum appears.  At the temperature $T_c$, the SUSY vacuum becomes the
true minimum and tunneling between the two vacua can occur.
We obtain the thermal bounce action, $B_\text{th} = 
S_\text{th}(\overline \phi(r)) - S_\text{th}(\phi_0)$, by solving 
the three dimensional versions of Eqs.~(\ref{eq:vacuum bounce},\ref{eq:boundary}) .

To evaluate the thermal tunneling rate $\Gamma \sim T^4
e^{-B_\text{th}/T}$, we make the following approximations.  We first
assume that the thermal bounce $B_\text{th}$ is a constant and
determine it numerically using the zero temperature potential, again
using the one-dimensional approximation.  We find that in order for
the universe to be stable, $B_\text{th}/T_c \gtrsim 230$.  The second
approximation we use is the high temperature expansion to determine
$T_c$.  We make this approximation because it is numerically difficult
to find when minima appear in a five dimensional space when the exact
one loop integrals do not converge quickly.  We find $T_c$ tends to be
smaller than $v$, where the high temperature expansion is not
valid.  Therefore, the result here should be viewed as only qualitatively
correct.

\begin{figure}[t]
  \centering
  \includegraphics[width=0.4\textwidth]{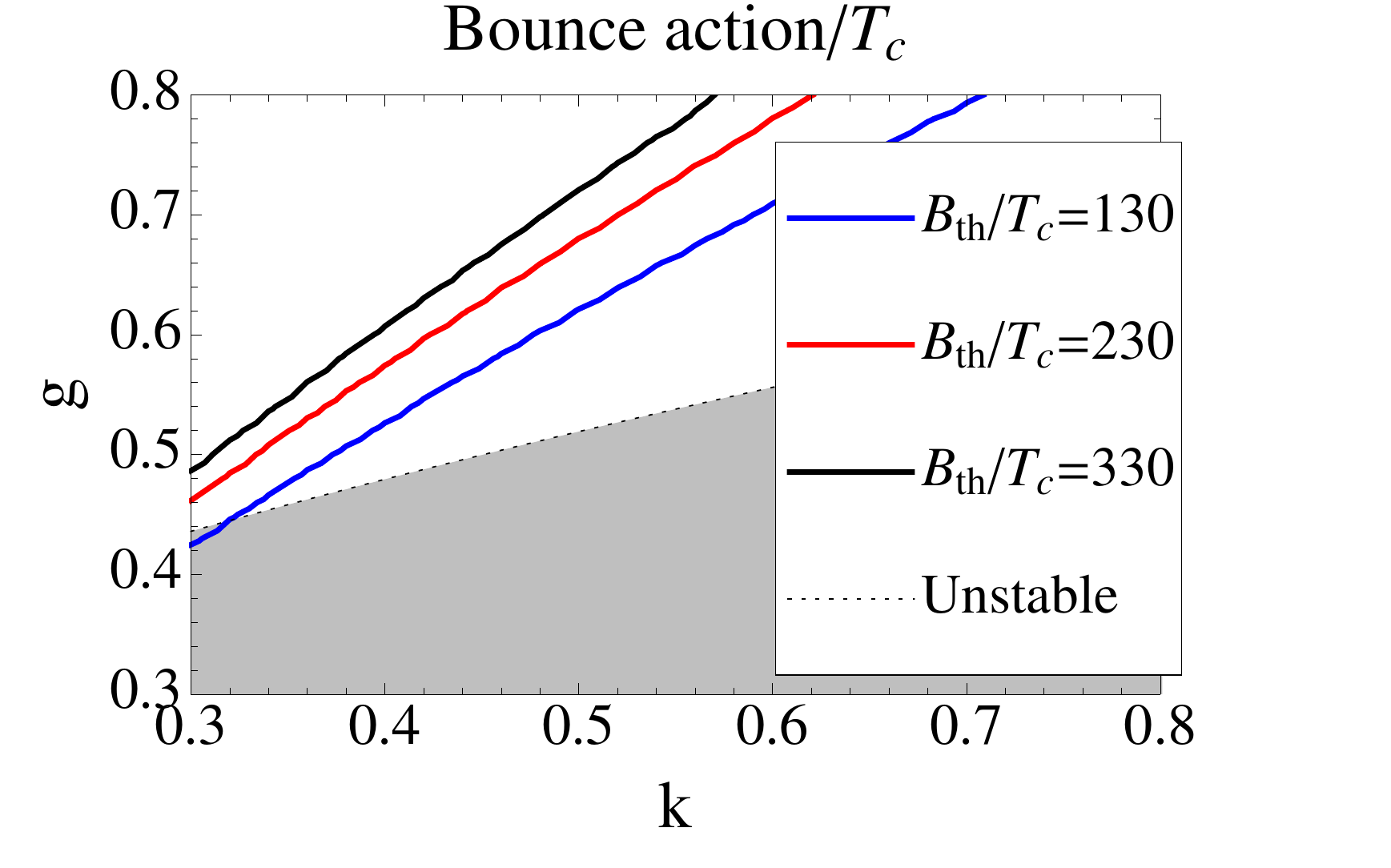}
  \caption{Contours of $B_\text{th}/T_c$ on the $(k,g)$ plane.  The
    solid blue, red, and black lines are for
    $B_\text{th}/T_c= 130$, 230, and 330, respectively.  
    For the universe to be stable, $B_\text{th}/T_c \gtrsim 230$.  The
    high temperature expansion is used to estimate $T_c$ and hence the
    result is meant to be only qualitatively correct.  See also
    the caption of Fig.~\protect\ref{fig:bounce}. }
  \label{fig:thermal bounce}
\end{figure}

We plot the contours of constant $B_\text{th}/T_c$ on the $(k,g)$
plane in Fig.~\ref{fig:thermal bounce}.  If the
value of $B_\text{th}/T_c$ were off by a factor of two, 
the limit on $k$ changes by 20\% for a fixed $g$.
Again, the one-dimensional approximation is poor in the shaded region
where messengers go tachyonic.

\paragraph{SUSY Spectrum}

All of the squarks and sleptons obtain the same mass-squared from the
non-zero $D$-term, 
\begin{equation}
  m^2_{\bf \tilde{\Psi}} = e D > 0 .
\end{equation}
and hence there is no flavor problem.

As mentioned earlier, gaugino masses are generated from one-loop
diagrams involving $\Phi$.  The mass of the fermion component of
$\Phi$ is $M_\Phi = g Y_0$, while the boson components have the mass
matrix (see Eq.~\eqref{Eq: messenger}),
\begin{equation}
  {\cal L} \supset -\frac{1}{2}
  (\tilde{\Phi}^*, \tilde{\Phi})
  \left(
    \begin{array}{cc}
      (g Y_0)^2 - 2e D & k F_S\\
      k F_S & (g Y_0)^2 - 2e D\\
    \end{array}
    \right)
    \left(
      \begin{array}{c}
        \tilde{\Phi} \\ \tilde{\Phi}^*
      \end{array}
    \right).
    \label{eq:massPhi}
\end{equation}
We choose
$g = 0.5$ and $k = 0.35$, in addition to the parameters in the
previous section.  There are no tachyons and the metastable
vacuum is stable on timescales of order the age of the universe.

At this point we need to discuss the Higgs.  In supersymmetric $E_6$
theories, we can regard the doublets in $\Phi$ as the Higgs
superfields.  We assign positive matter parity to doublets in one of
$\Phi$ so that it can have Yukawa couplings.  On the other hand, all
triplets and other doublets have negative matter parity so that they
do not have Yukawa couplings to $\Psi$.  They are simply messengers
for the gaugino mass.  The Higgses have the same mass matrix as in
Eq.~\eqref{eq:massPhi}, and we fine-tune $\mu = g_H Y_0$ against $2e
D$ to obtain a mass-squared at the 100~GeV scale.  It represents a
tuning at the level of $10^{-5}$.  $B\mu = k_H F_S$ can also be made
at the 100~GeV scale by choosing $k_H\approx 10^{-4}$.  Note that the
bare $\mu$-parameter is forbidden and is generated only with
$U(1)_\psi$ breaking, ameliorating the $\mu$-problem from the Planck
scale to the 100~TeV scale.

We have three color triplets and two electroweak doublets acting as
messengers to generate the gaugino masses.  The 1-loop results for the
gaugino masses can be found in Eq.~(2.3) of
Ref.~\cite{Poppitz:1996xw}.  
The gaugino masses can be enhanced relative to the standard gauge
mediation $3 \frac{\alpha_s}{4\pi} \frac{kF_S}{M_\Phi}$ due to the
$D$-term.  For the above numerical example, the enhancement factor is
1.39.  The on-shell gluino mass is further enhanced from $M_3$ by the
QCD correction by $1+\frac{4}{3} \frac{\alpha_s(M_3)}{\pi}$.

In these models, the scalars are heavy, the gauginos are light and the
gravitino is the Lightest Supersymmetric Particle.  For this type of mass spectrum,
the current ATLAS bound on the gluino mass is 1.28 TeV~\cite{ATLAS}
while the current CMS bound is around 1.15 TeV~\cite{CMS:2014koa}.
Because our gaugino masses are enhanced with respect to standard gauge
mediation, it is not very difficult to obtain a heavy enough gaugino.  For
the choice of parameters mentioned earlier, we find $v \geq 87$~TeV
from the gluino mass limit.

In the large $\lambda$ limit, $Y$ and $Z$ can be integrated out, and
the mass spectrum of the singlets can be computed as follows.  Among
the singlets, $S$ is a complex scalar and $X$ has only its real part
(the imaginary part is eaten by the $U(1)_\psi$ gauge boson), with masses
\begin{eqnarray}
  m_S^2 &=& M^2 \frac{X_0^2}{X_0^2 + Y_0^2} + 16 e^2 (-X_0^2 +
  Y_0^2), \\
  m_X^2 &=& M^2 \frac{X_0^2+ 16 e^2 (3X_0^2 +
  5Y_0^2)}{X_0^2 + Y_0^2}  \ .
\end{eqnarray}
Fermions of $X$ and $S$ become a Dirac particle of mass
\begin{equation}
  m_F^2 =  M^2 \frac{X_0^2}{X_0^2 + Y_0^2} 
    + 32 e^2 (X_0^2 + Y_0^2) .
\end{equation}
The $U(1)_\psi$ gauge boson acquires a mass
\begin{equation}
  m_V^2 = 32 e^2 (X_0^2 + Y_0^2).
\end{equation}
We have already presented the mass spectrum of the $\Phi$ and $\Psi$
multiplets.  We verified that ${\rm Str} M^2 = 0$ once spectrum
integrated out at $O(\lambda^2 (X_0^2 + Y_0^2))$ is included. 

\begin{figure}[t]
  \centering
  \includegraphics[width=0.4\textwidth]{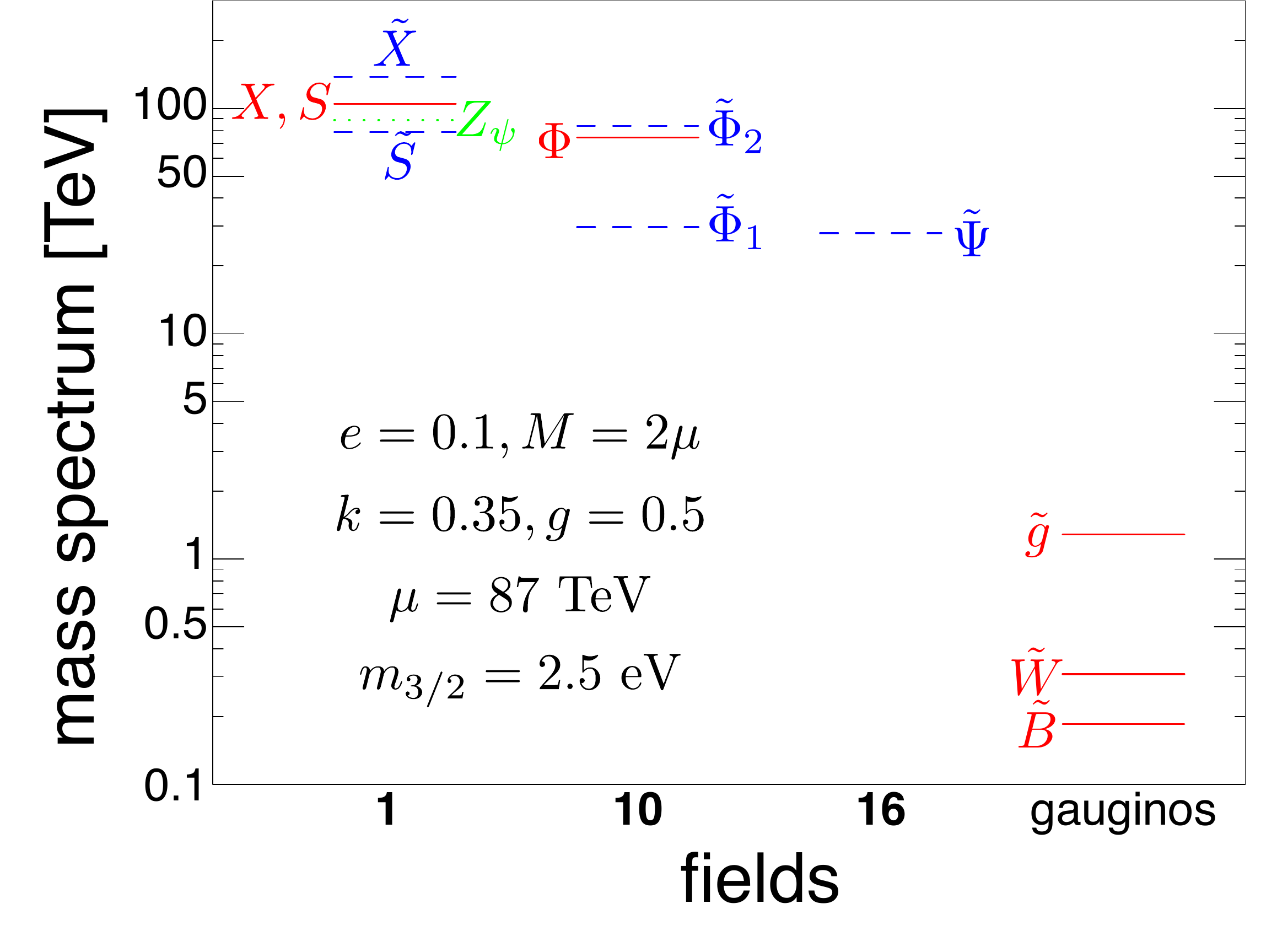}
  \caption{An example mass spectrum in our model.  The fields are
    presented according to the $SO(10)$ representations, {\bf 1}, {\bf
      10}, and {\bf 16}, as well as the Standard Model gauginos.  The
    gravitino is too light to be shown.}
  \label{fig:spectrum}
\end{figure}

In order for the gluino to be heavier than a TeV, we typically have
scalars at 20--100~TeV.  If $e \sim O(1)$, this is a similar mass
spectrum to mini-split~\cite{Arvanitaki:2012ps}, pure gravity
mediation~\cite{Ibe:2011aa}, and SUSY breaking mechanisms with
accidental $R$-symmetries, {\it e.g.}\/, metastable
Intriligator-Seiberg-Shih (ISS) SUSY
breaking~\cite{Intriligator:2006dd}.  The main difference of vector
mediation with previous models is the presence of a superlight
gravitino.  The mass spectrum is shown in Fig.~\ref{fig:spectrum}.
The radiative corrections from stops at the 10~TeV scale easily raise
the lightest Higgs boson mass to the observed mass of 125~GeV.  Note
that the fine-tuning is smaller compared to other models with a split
spectrum because the mass of three {\bf 16} is at $\sim 30$~TeV whereas the
other scalar masses are at $\sim 100$~TeV.  This is because the $U(1)_\psi$
charges differ by a factor of four and the $D$-term is always smaller
than the $F$-term.  These scalars may be discovered at a 100~TeV $pp$
collider.

\paragraph{Implications}

The most important implication is that the gravitino mass is
\begin{equation}
  m_{3/2}^2 = \frac{V}{3M_{Pl}^2}
  = \frac{1.9 v^4}{3 M_{Pl}^2}
  = (2.5~{\rm eV})^2
\end{equation}
for the above parameters.  Here, $M_{Pl}=2.4 \times 10^{18}$~GeV.
Such a low mass gravitino is not possible in the usual low-energy
gauge mediation and is completely compatible with the cosmological
limit of $m_{3/2} < 16$~eV. There is a clear upper limit on the gluino
mass from the cosmology of about 5~TeV.  The LHC would most likely
discover the gluino, and a 100~TeV $pp$ collider would completely
close the window.  The bino decays promptly $\tilde{B} \rightarrow
\gamma \tilde{G}$, giving evidence of low-energy SUSY breaking.  The ratio
among gaugino masses is also unique, different from the standard
gauge-mediation or unification scenarios.



The right-handed neutrino is charged under $U(1)_\psi$ and hence we
cannot use the seesaw mechanism for neutrino masses.  One can use a
linear combination of $U(1)_\psi$ and $U(1)_\chi$ instead to make the
right-handed neutrino neutral \footnote{HM thanks Yanagida on this
  point.}.  This is an interesting variation to be considered.  In the
simplest realization of these models presented in this Letter, the
right-handed neutrinos are Dirac, are present at low energies, and
can contribute to energy densities at BBN.  Much of the number density
in right handed neutrinos is diluted away by the entropy generated at
the QCD phase transition, resulting in $\Delta N_\nu \simeq 0.19$.  This
may be verified in the future cosmological data.


HM thanks T.T. Yanagida for discussions.  We thank the Aspen Center
for Physics and the NSF Grant \#1066293 for hospitality during the
Aspen Winter Conference ``{\it Exploring the Physics Frontier with
  Circular Colliders}\/'' where our collaboration began.  The work of
AH is supported by the DOE grant DE-SC0009988.  HM is supported in
part by the U.S. DOE under Contract DE-AC03-76SF00098, in part by the
NSF under grant PHY-1002399, in part by the JSPS Grant-in-Aid for
Scientific Research (C) (No. 26400241), Scientific Research on
Innovative Areas (No. 26105507), and by WPI, MEXT, Japan.

\bibliography{refs}
\end{document}